\documentclass[12pt]{article}

\usepackage{graphicx}

\usepackage{hyperref}

\def\beq{\begin{equation}}
\def\eeq{\end{equation}}
\def\bea{\begin{eqnarray}}
\def\eea{\end{eqnarray}}
\def\bwt{\begin{widetext}}
\def\ewt{\end{widetext}}
\def\nn{\nonumber}
\def\roughly#1{\mathrel{\raise.3ex\hbox
{$#1$\kern-.75em\lower1ex\hbox{$\sim$}}}}
\def\lsim{\roughly<}
\def\gsim{\roughly>}

\def\bd{B^0}

\def\order{\lower 1.8ex \hbox{\LARGE\~{}}}

\def\BDtaunu{\bar{B} \to D^+ \tau^{-} {\bar\nu}_\tau}
\def\BDlnu{\bar{B} \to D^+ \ell^{-} {\bar\nu}_\ell}
\def\BDstartaunu{\bar{B} \to D^{*+} \tau^{-} {\bar\nu}_\tau}
\def\BDstarlnu{\bar{B} \to D^{*+} \ell^{-} {\bar\nu}_\ell}
\newcommand{\bctaunui}{b \to c \tau^- {\bar\nu}_i}

\def \kstar{K^*}
\def\BKstarmumu{\bd \to \kstar \mu^+ \mu^-}

\def \({\left(}
\def \){\right)}
\def \[{\left[}
\def \]{\right]}
\def \l|{\left|}
\def \r|{\right|}

\def \Re{\rm Re}

\def \nn{\nonumber}
\def \nl{\nn \\}

\def \al{\alpha}

\def \ga{\gamma}
\def \Ga{\Gamma}
\def \de{\delta}

\def \la{\lambda}
\def \La{\Lambda}

\def \ka{\kappa}

\def \SM{{\rm SM}}
\def \EM{{\rm EM}}
\def \NP{{\rm NP}}

\def \kSM{\ka^{}_\SM}
\def \aEM{\al^{}_\EM}
\def \GF{G^{}_F}

\def \s{\sqrt{2}}

\def \btos{b \to s}
\def \bbtosb{{\bar b}\to{\bar s}}

\begin{document}

\begin{flushright}
UdeM-GPP-TH-14-239 \\
UMISS-HEP-2014-03 \\
\end{flushright}

\begin{center}
\bigskip
{\Large \bf \boldmath Simultaneous Explanation of the \\ $R_K$ and $R(D^{(*)})$ Puzzles} \\
\bigskip
\bigskip
{\large
Bhubanjyoti Bhattacharya $^{a,}$\footnote{bhujyo@lps.umontreal.ca},
Alakabha Datta $^{b,}$\footnote{datta@phy.olemiss.edu}, \\
David London $^{a,}$\footnote{london@lps.umontreal.ca}
and Shanmuka Shivashankara $^{b,}$\footnote{sshivash@go.olemiss.edu}
}
\end{center}

\begin{flushleft}
~~~~~~~~~~~$a$: {\it Physique des Particules, Universit\'e
de Montr\'eal,}\\
~~~~~~~~~~~~~~~{\it C.P. 6128, succ. centre-ville, Montr\'eal, QC,
Canada H3C 3J7}\\
~~~~~~~~~~~$b$: {\it Department of Physics and
  Astronomy, 108 Lewis Hall, }\\
~~~~~~~~~~~~~~~{\it University of Mississippi, Oxford, MS
  38677-1848, USA }
\end{flushleft}

\begin{center}
\bigskip (\today)
\vskip0.5cm {\Large Abstract\\} \vskip3truemm
\parbox[t]{\textwidth}{At present, there are several hints of
  lepton flavor non-universality.  The LHCb Collaboration has measured $R_K
  \equiv {\cal B}(B^+ \to K^+ \mu^+ \mu^-)/{\cal B}(B^+ \to K^+ e^+
  e^-)$, and the BaBar Collaboration has measured $R(D^{(*)}) \equiv
  {\cal B}({\bar B} \to D^{(*)+} \tau^- {\bar\nu}_\tau) / {\cal
    B}({\bar B} \to D^{(*)+} \ell^- {\bar\nu}_\ell)$ ($\ell =
  e,\mu$). In all cases, the experimental results differ from the
  standard model predictions by 2-3$\sigma$.  Recently, an explanation
  of the $R_K$ puzzle was proposed in which new physics (NP) generates
  a neutral-current operator involving only third-generation
  particles. Now, assuming the scale of NP is much larger than the
  weak scale, this NP operator must be made invariant under the full
  $SU(3)_C \times SU(2)_L \times U(1)_Y$ gauge group.  In this Letter,
  we note that, when this is done, a new charged-current operator can
  appear, and this can explain the $R(D^{(*)})$ puzzle. A more precise
  measurement of the double ratio $R(D)/R(D^*)$ can rule out this
  model.}

\end{center}

\thispagestyle{empty}
\newpage
\setcounter{page}{1}
\baselineskip=14pt

To date, the standard model (SM) has been extremely successful in
describing experimental data. There are, however, a few measurements
that are in disagreement with the predictions of the SM. For example,
the LHCb Collaboration recently measured the ratio of decay rates
for $B^+ \to K^+ \ell^+ \ell^-$ ($\ell = e,\mu$) in the dilepton
invariant mass-squared range 1 GeV$^2$ $\le q^2 \le 6$ GeV$^2$
\cite{RKexpt}. They found
\beq
R_K \equiv \frac{{\cal B}(B^+ \to K^+ \mu^+ \mu^-)}{{\cal B}(B^+ \to
  K^+ e^+ e^-)}
= 0.745^{+0.090}_{-0.074}~{\rm (stat)} \pm 0.036~{\rm (syst)} ~,
\eeq
which is a $2.6\sigma$ difference from the SM prediction of $R_K = 1
\pm O(10^{-4})$ \cite{RKtheory}. As another example, the BaBar
Collaboration with their full data sample has reported the
following measurements \cite{RDexpt1,RDexpt2}:
\begin{eqnarray}
\label{babarnew}
R(D) &\equiv& \frac{{\cal B}(\BDtaunu)}
{{\cal B}(\BDlnu)}=0.440 \pm 0.058 \pm 0.042 ~, \nn \\
R(D^*) &\equiv& \frac{{\cal B}(\BDstartaunu)}
{{\cal B}(\BDstarlnu)} = 0.332 \pm 0.024 \pm 0.018 ~,
\label{RDexpt}
\end{eqnarray}
where $\ell = e,\mu$. The SM predictions are $R(D) = 0.297 \pm 0.017$
and $R(D^*) = 0.252 \pm 0.003$ \cite{RDexpt1,RDtheory}, which deviate
from the BaBar measurements by 2$\sigma$ and 2.7$\sigma$,
respectively. (The BaBar Collaboration itself reported a 3.4$\sigma$
deviation from SM when the two measurements of Eq.~(\ref{babarnew})
are taken together.) These two measurements of lepton flavor non-universality,
respectively referred to as the $R_K$ and $R(D^{(*)})$
puzzles, may be providing a hint of the new physics (NP) believed to
exist beyond the SM.

In addition, we note that the three-body decay $\BKstarmumu$ by itself
offers a large number of observables in the kinematic and angular
distributions of the final-state particles, and it has been argued
that some of these distributions are less affected by hadronic
uncertainties \cite{BKmumuhadunc}. Interestingly, the measurement of
one of these observables shows a deviation from the SM prediction
\cite{Aaij:2013qta}. However, the situation is not clear whether this
anomaly is truly a first sign of new physics. There are unknown
hadronic uncertainties that must be taken into account before one can
draw this conclusion \cite{BKmumuNP,HurthAS,bsnunubar}. We therefore do
not discuss this measurement further.

To search for an explanation of $R_K$, in Ref.~\cite{HS1} Hiller and
Schmaltz perform a model-independent analysis of $\btos \ell^+ \ell^-$.
They consider NP operators of the form $({\bar s} {\cal O} b)({\bar \ell}
{\cal O}' \ell)$, where ${\cal O}$ and ${\cal O}'$ span all Lorentz
structures. They find that the only NP operator that can reproduce the
experimental value of $R_K$ is $({\bar s}\gamma_\mu P_L b)({\bar \ell}
\gamma^\mu P_L \ell)$. This is consistent with the NP explanations for
the $B\to K^{(*)}\mu^+\mu^-$ angular distributions measured by LHCb
\cite{HurthAS}.

In Ref.~\cite{GGL}, Glashow, Guadagnoli and Lane (GGL) note that
lepton flavor non-universality is necessarily associated with lepton flavor
violation (LFV). With this in mind, they assume that the NP couples
preferentially to the third generation, giving rise to the operator
\beq
G ({\bar b}'_L \gamma_\mu b'_L) ({\bar \tau}'_L \gamma^\mu \tau'_L) ~,
\label{GGLoperator}
\eeq
where $G = O(1)/\Lambda_{NP}^2 \ll G_F$, and the primed fields are the
fermion eigenstates in the gauge basis. The gauge eigenstates are
related to the physical mass eigenstates by unitary transformations
involving $U_L^d$ and $U_L^\ell$:
\beq
d'_{L3} \equiv b'_L = \sum_{i=1}^3 U^d_{L3i} d_i ~~,~~~~
\ell'_{L3} \equiv \tau'_L = \sum_{i=1}^3 U^\ell_{L3i} \ell_i ~.
\eeq
With this, Eq.~(\ref{GGLoperator}) generates an NP operator that
contributes to $\bbtosb \mu^+ \mu^-$:
\beq
G \left[ U^d_{L33} U^{d*}_{L32} |U^\ell_{L32}|^2 ({\bar b}_L
\gamma_\mu s_L) ({\bar \mu}_L \gamma^\mu \mu_L) + h.c. \right].
\label{GGLoperator2}
\eeq

Because the coefficient of this operator involves elements of the
mixing matrices, which are unknown, one cannot make a precise
evaluation of the effect of this operator on ${\cal B}(B^+ \to K^+
\mu^+ \mu^-)$, and hence on $R_K$. Still, GGL note that the hierarchy
of the elements of Cabibbo-Kobayashi-Maskawa quark mixing matrix,
along with the apparent preference of the NP for muons over electrons,
suggests that $|U^{d,\ell}_{L33}| \simeq 1$ and $|U^{d,\ell}_{L31}|^2
\ll |U^{d,\ell}_{L32}|^2 \ll 1$. Furthermore, there are limits on some
ratios of magnitudes of matrix elements. Taken together, GGL find that
the observed value of $R_K$ can be accommodated with the addition of
the NP operator in Eq.~(\ref{GGLoperator2}).

In any case, GGL's main point is not so much to offer
Eq.~(\ref{GGLoperator}) as an explanation of $R_K$, but rather to
stress that the NP responsible for the lepton flavor non-universality will
generally also lead to an enhancement of the rates for
lepton-flavor-violating processes such as $B\to K \mu e, K \mu \tau$
and $B_s \to \mu e, \mu \tau$. In the case of Eq.~(\ref{GGLoperator}),
it is clear how LFV arises. This operator is written in terms of the
fermion fields in the gauge basis and does not respect lepton-flavor
universality. In transforming to the mass basis, the GIM mechanism
\cite{GIM} is broken, and processes with LFV are generated.

In fact, this behavior is quite general. In writing down effective
Lagrangians, it is usually only required that the operators respect
$SU(3)_C \times U(1)_{em}$ gauge invariance. However, it was argued in
Refs.~\cite{HS1,AGC} that if the scale of NP is much larger than the
weak scale, the operators generated when one integrates out the NP
must be invariant under the full $SU(3)_C \times SU(2)_L \times
U(1)_Y$ gauge group. In the same vein, the operators should be written
in terms of the fermion fields in the gauge basis -- after all, above
the weak scale, the mass eigenstates do not (yet) exist. If these
operators break lepton universality, lepton-flavor-violating
interactions will appear at low energy when one transforms to the mass
basis. (Note, however, that in explicit models one can avoid lepton flavor
non-universality and lepton flavor violation through the imposition of
additional symmetries. One such example can be found in Ref.\
\cite{Altmanshoffer}.)

There have been a number of analyses, both model-independent and
model-dependent, examining explanations of the $R_K$ puzzle.
(Sometimes the data from the $B\to K^{(*)}\mu^+\mu^-$ angular
distributions were also included.) In all cases, the low-energy
operators were written in terms of mass eigenstates, and
lepton-flavor-violating operators were not included.  However, as
argued above, such operators will appear when lepton universality is
broken. Now, the model-independent analyses
\cite{HurthAS,HS1,AGC,modelindRK} will be little changed by the inclusion
of such operators. However, considerations of such
lepton-flavor-violating interactions would be useful in the context of
model-dependent analyses. Leptoquarks \cite{HS1, GNR} and
R-parity-violating SUSY \cite{RPVSUSY} have been proposed as possible
solutions to the $R_K$ puzzle. In both cases, it would be interesting
to examine the predictions for the lepton-flavor-violating processes.

Coming back to the GGL operator of Eq.~(\ref{GGLoperator}), it too
must be made invariant under $SU(3)_C \times SU(2)_L \times U(1)_Y$.
There are two consequences. First, the left-handed fermion fields must
be replaced by $SU(2)_L$ doublets: $b'_L \to Q'_L$ and $\tau'_L \to
L'_L$, where $Q' \equiv (t',b')^T$ and $L' \equiv
(\nu'_\tau,\tau')^T$.  Second, there are two NP operators that are
invariant under $SU(2)_L$ and contain Eq.~(\ref{GGLoperator}):
\bea
{\cal O}^{(1)}_{NP} &=& G_1 ({\bar Q}'_L \gamma_\mu Q'_L) ({\bar L}'_L \gamma^\mu L'_L) ~, \nn\\
{\cal O}^{(2)}_{NP} &=& G_2 ({\bar Q}'_L \gamma_\mu \sigma^I Q'_L) ({\bar L}'_L \gamma^\mu \sigma^I L'_L) ~,
\eea
where $G_1$ and $G_2$ are both $O(1)/\Lambda_{NP}^2$ (but not equal to
one another), and $\sigma^I$ are the Pauli matrices (the generators of
$SU(2)$).  Using the identity
\beq
\sigma^{I}_{ij}\sigma^{I}_{kl} = 2 \delta_{il}\delta_{kj} - \delta_{ij}\delta_{kl} ~,
\eeq
where $i, j$ are $SU(2)_L$ indices, the second operator can be
written as
\beq
{\cal O}^{(2)}_{NP} = G_2 \left[
2 ({\bar Q}'^{i}_L \gamma_\mu Q'^{j}_L) ({\bar L}'^{j}_L \gamma^\mu L'^{i}_L)
- ({\bar Q}'_L \gamma_\mu Q'_L) ({\bar L}'_L \gamma^\mu L'_L) \right] ~.
\eeq
The two operators correspond to different types of underlying NP.
Specifically, ${\cal O}^{(1)}_{NP}$ contains only neutral-current (NC)
interactions, while ${\cal O}^{(2)}_{NP}$ contains both
neutral-current and charged-current (CC) interactions.  ${\cal
  O}^{(2)}_{NP}$ therefore offers the potential to simultaneously
explain both the $R_K$ and $R(D^{(*)})$ puzzles, and we examine the
effects of including this NP operator.

Writing ${\cal O}^{(2)}_{NP}$ explicitly in terms of the up-type and
down-type fields, there are four NC operators and one CC operator:
\bea
{\cal O}^{(2)}_{NP} & = & O_{ t t \nu_\tau \nu_\tau} + O_{ b b \tau \tau} +
O_{ t t \tau \tau} + O_{ b b \nu_\tau \nu_\tau} + O_{t b \tau \nu_\tau} ~,
\label{explicit}
\eea
with
\bea
O_{ t t \nu_\tau \nu_\tau} & = & G_2 ({\bar t}'_L \gamma_\mu t'_L) ({\bar \nu}'_{\tau_L} \gamma^\mu \nu'_{\tau_L}) ~, \nn\\
O_{ b b \tau \tau} & = & G_2 ({\bar b}'_L \gamma_\mu b'_L) ({\bar\tau}'_L \gamma^\mu \tau'_L) ~, \nn\\
O_{ t t \tau \tau} & = & -G_2 ({\bar t}'_L \gamma_\mu t'_L) ({\bar\tau}'_L \gamma^\mu \tau'_L) ~, \nn\\
O_{ b b \nu_\tau \nu_\tau} & = & -G_2 ({\bar b}'_L \gamma_\mu b'_L) ({\bar\nu}'_{\tau L} \gamma^\mu \nu'_{\tau L}) ~, \nn\\
O_{t b \tau \nu_\tau} &=& 2 G_2 ({\bar t}'_L \gamma_\mu b'_L) ({\bar \tau}'_L \gamma^\mu \nu'_{\tau L}) + h.c.
\eea
If both ${\cal O}^{(1)}_{NP}$ and ${\cal O}^{(2)}_{NP}$ are present
then the NC interactions receive contributions from both NP
operators.

Above, we see that the NC part of ${\cal O}^{(2)}_{NP}$ contains
$O_{b b \tau \tau}$, which is the GGL operator of Eq.~(\ref{GGLoperator}).
In transforming to the mass basis, the GGL piece therefore contributes
to $\bbtosb$ transitions through the quark-level decays $\bbtosb \ell^+
\ell^-$ and $\bbtosb \ell^+\ell^{\prime -}$. These generate the meson-level
decays $B \to K^{(*)}\mu^+ \mu^-, B \to K^{(*)} \mu^\pm e^\mp, B \to K^{(*)}
\mu^\pm\tau^\mp, B_s \to \mu^+ \mu^-, B^0\to X_s \mu^+ \mu^-, B^0_s \to \mu^+
\mu^- \gamma,$ etc.  (Many of these decays are discussed by GGL.) The largest
effects will be an enhancement of the SM contribution to $\bbtosb\tau^+ \tau^-$,
and the generation of the lepton-flavor-violating decays $\bbtosb \tau^{\pm}
\mu^{\mp}$\cite{Boubaa:2012xj}.

We begin by discussing the effect of ${\cal O}^{(2)}_{NP}$ on $R_K$. The
amplitude for $\bbtosb \ell^+_i\ell^-_i$ ($\ell_1 = e, \ell_2 = \mu$) can
be expressed as
\beq
A^{}_{\ell_i} ~=~ A^{\SM}\(1 + V^{bs\ell_i}_L\)~,~~~~ V^{bs\ell_i}_L ~=~
\frac{\ka}{C^{}_9}\frac{U^{d}_{L33} U^{d*}_{L32}}{V^{}_{tb}V^*_{ts}}
|U^{\ell}_{L3i}|^2~,~~~~ \ka ~=~ \frac{4\pi}{\aEM}\frac{g^2_2}{g^2}\frac{M^2_W}
{\La^2_\NP}~.
\eeq
Here $A^{\SM}$ is the lepton-flavor-universal (SM) contribution, the
$V_{ij}$ are Cabibbo-Kobayashi-Maskawa (CKM) matrix elements, $C^{}_9$
is a Wilson coefficient, and we have written $G^{}_2 = g^2_2/\La^2_\NP$.
Neglecting the masses of the leptons we then arrive at the following result:
\bea
R^{}_K &=& \frac{1 + 2 \, {\Re}[V^{bs\mu}_L] + |V^{bs\mu}_L|^2}{1 + 2 \, {\Re}
[V^{bse}_L] + |V^{bse}_L|^2} \nl
&\approx& 1 + \frac{8\pi}{C^{}_9\aEM}\frac{g^2_2}{g^2}\frac{M^2_W}{\La^2_\NP}
\frac{U^{d}_{L32}|U^{\ell}_{L32}|^2}{\la^2}~,~~
\eea
where $\la$ is the sine of the Cabibbo angle. We have assumed the
usual hierarchy of CKM matrix elements and ignored all CP-violating
phases. The $5\sigma$ limit on $R^{}_K$ from LHCb then implies
\bea
- 2\times10^{-4} \lsim \frac{1}{C^{}_9}\frac{g^2_2}{g^2}\frac{M^2_W}{\La^2_\NP}
\frac{U^{d}_{L32}|U^{\ell}_{L32}|^2}{\la^2} \lsim 7\times 10^{-5}~.~~
\label{con1}
\eea
It is clear that the LHCb measurement constrains the magnitudes of the down-type
and lepton mixing-matrix elements. However, a further set of constraints will be
obtained below.

In addition to the decays produced by the GGL operator, one now also
has the quark-level decay $\bbtosb \nu {\bar\nu}$ that contributes to
$B \to K^{(*)} \nu {\bar\nu}$. The amplitude for $\bbtosb \nu^{}_i
{\bar\nu}^{}_j$ can be expressed as
\bea
A^{}_{ij} &=& C^{}_{ij} ({\bar b}_L \ga^{}_\mu s_L)
({\bar\nu}_{iL} \ga^\mu \nu_{jL})~.
\eea
The SM contributes only to terms diagonal in neutrino flavor ($i=j$),
while the NP operator also gives rise to off-diagonal terms that
violate lepton flavor ($i\ne j$). We have
\bea
C^{}_{ij} &=& \kSM \(\de^{}_{ij} - \frac{\ka}{C^{\SM}_L}\frac{U^{d}_{L33}
U^{d*}_{L32}}{V^{}_{tb}V^*_{ts}}U^{*\nu}_{L3i}U^\nu_{L3j}\)~,~~
\eea
where
\bea
\kSM &=& \frac{\s\GF\aEM}{\pi}V^{}_{tb}V^*_{ts} C^\SM_L~.~~ 
\eea
In the above, $C^\SM_L$ is a Wilson coefficient \cite{bsnunubar}.
The square of the amplitude for the process is thus proportional to
\bea
\sum\limits_{i, j} |C^{}_{ij}|^2 &=& 3|\kSM|^2\(1 - \frac{2\ka}{3}
{\Re}\[x\] + \frac{\ka^2}{3}|x|^2\)~,~~
\eea
where $x = \(U^{d}_{L33}U^{d*}_{L32}\)/\(V^{}_{tb}V^*_{ts}\)$.

We ignore all CP-violating phases, so that $x$ is real. Taking
$|U^{d}_{L33}|\sim1$, we have $x \sim U^{d}_{L32}/\la^2$.
The decay rate for $B \to K^{(*)}\nu {\bar\nu}$ is given by
\bea
\Ga &=& \Ga^{}_\SM\(1 - \frac{2\ka U^d_{L32}}{3\la^2} + \frac{(\ka U^d_{L32})^2}
{3\la^4}\) ~.~~
\eea
The SM decay rate can be expressed as follows:
\bea
\Ga^{}_\SM &=& \frac{m^{}_B|\kSM|^2}{64\pi^3}\int\limits^{q^2|_{\rm max}}_{0}
\rho^{}_{K^{(*)}}(q^2)dq^2~,~~
\eea
where $q$ represents the sum of four momenta of the neutrino and the
antineutrino, and $\rho^{}_{K^{(*)}}$ is the appropriate $B\to
K^{(*)}$ transition form factor. (Note that we have treated the
neutrinos as massless particles.) Thus we see that the NP term simply
modifies the SM rate for $B\to K\nu{\bar\nu}$ by an overall numerical
factor.

One can use the above result to get an estimate of how large the NP
couplings and mixing matrix elements can be. A precise calculation
of the SM branching ratio for $B^+\to K^+\nu{\bar \nu}$ was performed
in Ref.~\cite{bsnunubar}. It was found that
\bea
{\cal B}(B^+\to K^+\nu{\bar\nu})_{\SM} &=& (4.20\pm0.33\pm0.15)\times10^{-6}~.
\eea
The strongest experimental bounds from the BaBar Collaboration
\cite{bsnunuex} at present only set an upper limit of $1.7\times
10^{-5}$ at the $90\%$ confidence level. Thus there is still room for
the measured decay rate to be a factor of five larger than the SM
prediction. Taking $C^\SM_L \approx -6.13$ \cite{bsnunubar}, we have
$\ka \simeq - 281 (g^{}_2/g)^2(M^{}_W/\La_\NP)^2 $. A factor of five
enhancement in the decay rate due to the NP operator ${\cal O}^{(2)}_{NP}$
would then imply
\bea
- 1.6\times 10^{-2} \lsim \frac{g^2_2}{g^2}\frac{M^2_W}{\La^2_\NP}
\frac{U^d_{L32}}{\la^2} \lsim 9.3\times 10^{-3} ~.~~ \label{con2}
\eea
If $\La^{}_\NP \simeq 10 M^{}_W$ then $(g^2_2/g^2)(U^d_{L32}/\la^2)$
must be $O(1)$.  In this case, a NP coupling of the same order as that
of the SM will still allow a reasonably large value for $U^d_{L32}$.
For example, if $g/2 \lsim g_2 \lsim g$, one can have $\la \gsim
U^d_{L32} \gsim \lambda^2$. In addition, we can now combine
Eqs.~(\ref{con1}) and (\ref{con2}). Since $C^{}_9$ is an $O(1)$
number, this implies that an $O(10^{-1})$ value for $|U^{l}_{L32}|$ is
still allowed. A more precise measurement of both $R^{}_K$ and $B^+\to
K^+ \nu{\bar \nu}$ will put stricter bounds on both the down-type and
lepton mixing-matrix elements.

Finally, the neutral-current part of ${\cal O}^{(2)}_{NP}$ also
contributes to the decays $t \to c \ell^+ \ell^-$, $t \to c \ell^+
\ell^{\prime -}$ and $t \to c \nu {\bar\nu}$.  The branching ratios
for these decays are negligible in the SM, so any observation would be
a clear sign of NP. For decays to charged leptons, the most promising
is $t \to c \tau^+ \tau^-$. In the mass basis, the contributing NP
operator is
\beq
G \[ U^{u^*}_{L32} \, U^u_{L33} \, |U^\ell_{L33}|^2 \, ({\bar c}_L \gamma^\mu
t_L)({\bar \tau}_L \gamma_\mu \tau_L) + h.c.\] ~,
\eeq
which gives a partial width of
\beq
\frac{g_2^4 |U^u_{L32}|^2 \, |U^u_{L33}|^2 \, |U^\ell_{L33}|^4 }{16
\Lambda_{NP}^4} \frac{m_t^5}{48 \pi^3} ~.
\eeq
Taking $g_2 \sim g $, $|U^u_{L33}| \simeq |U^\ell_{L33}| \simeq 1$,
$|U^u_{L32}| \simeq \lambda$, and $\Lambda_{NP} = 800$ GeV, this gives
\beq
\Gamma(t \to c \tau^+ \tau^-) = 1 \times 10^{-7}~~{\rm GeV}~.
\eeq
The full width of the $t$ quark is 2 GeV, so this corresponds to a
branching ratio of $5 \times 10^{-8}$. This is much larger than the
SM branching ratio ($O(10^{-16})$), but is still tiny. The branching
ratio for $t \to c \nu {\bar\nu}$ takes the same value, while those
for all other $t \to c \ell^+ \ell^-$ and $t \to c \ell^+ \ell^{\prime
  -}$ decays are considerably smaller. Thus, while the branching
ratios for these decays can be enormously enhanced compared to the SM,
they are still probably unmeasurable. (This point is also noted in
Ref.~\cite{HS1}.)

Another process involving $t$ quarks that could potentially reveal the
presence of NP with LFV is $pp \to t {\bar t}$, followed by the
radiation of a $\tau^\pm \mu^\mp$ pair. At the LHC with a 13 TeV
center-of-mass energy, gluon fusion dominates the production of
$t{\bar t}$ pairs.  We use MadGraph 5 \cite{madgraph5} to calculate
the cross section for $gg \to t {\bar t} \tau^\pm \mu^\mp$, taking
$g_2 \sim g$. We find $\sigma_{t{\bar t}\tau\mu} \approx 0.4
|U^\ell_{L32}|^{2}$ fb.  By contrast, the SM cross section for $t{\bar
  t}$ pair production is $\sigma_{t{\bar t}} \approx 450$ pb, so that
$\sigma_{t{\bar t}\tau\mu} / \sigma_{t{\bar t}} \approx 10^{-6}
|U^\ell_{L32}|^{2}$, which is extremely small.  With a luminosity of
100 $fb^{-1} $ /year at the 13 TeV LHC \cite{ttprod}, we therefore
expect about 40 events/year for $gg \to t {\bar t} \tau^\pm \mu^\mp$
if $|U^\ell_{L32}| \sim 1$, or about two events/year if $ |U^\ell_{L32}|
\sim \lambda $. Thus, even though the final-state signal is striking, $pp
\to t {\bar t} \tau^\pm \mu^\mp$ is probably unobservable.

Turning to the charged-current interactions, these contribute to both
$b$ and $t$ semileptonic decays. Even with the enhancement from NP,
the decay $t \to b \tau \bar{\nu_{\tau}}$ will still be difficult to
observe, as it is swamped by the two-body decay $t \to b W$.  On the
other hand, the decay $b \to c \tau {\bar\nu}_i$ ($i=\tau,\mu,e$) is
particularly interesting, since it contributes to the decay ${\bar B}
\to D^{(*)+} \tau^- {\bar\nu}_\tau$ and the $R(D^{(*)})$ puzzle
[Eq.~(\ref{RDexpt})], and provides a aource of lepton flavor non-universality
in such decays.

In the SM, the effective Hamiltonian for the quark-level transition $b
\to c \tau {\bar\nu}_\tau$ is
\beq
{\cal{H}}_{eff} = \frac{4 G_F V_{cb}}{\sqrt{2}} (\bar{c}_L \gamma_\mu b_L)
(\bar{\tau}_L \gamma^\mu \nu_{\tau L}) + h.c. ~.
\eeq
Now, if ${\cal O}^{(2)}_{NP}$ is also present, in addition to $\tau
{\bar\nu}_\tau$ in the final state, the NP operator also produces
$\tau {\bar\nu}_\mu$ and $\tau {\bar\nu}_e$.  However, as the
final-state neutrino is not observed, we have to sum over the neutrino
species. That is, the squared-amplitude for $\bctaunui$ can be written
as
\beq
|A|^2 = \sum_{i=\tau,\mu,e} |A^{}_i|^2 ~,
\eeq
with
\bea
A_i & = & \frac{4 G_F V_{cb}}{\sqrt{2}} \left[\de_{i\tau}
+ V_L^{cb\tau\nu_i}\right] ~~,~~~~
V_L^{cb\tau\nu_i} = 4 \, \frac{g_2^2}{g^2}\frac{M_W^2}{\Lambda_{NP}^2}
\frac{U^d_{L33} U^u_{L32} U^\ell_{L33} U_{L3i}^\nu}{V_{cb}} ~.
\label{NPbcl}
\eea
As was done above, we have written $G_2 \equiv g_2^2/\Lambda_{NP}^2$
and used $G_F/\sqrt{2}= g^2/8 M_W^2$. One then has
\beq
|A|^2 = |A|^2_{SM} \left[ 1+ 2 \, {\rm Re}(V_L^{cb\tau\nu_\tau}) + |V_L^{cb\tau}|^2 \right] ~,
\label{NPrate}
\eeq
where
\beq
|V_L^{cb\tau}|^2 \equiv \sum_i |V_L^{cb\tau\nu_i}|^2
= \left\vert 4 \, \frac{ g_2^2}{g^2}\frac{M_W^2}{\Lambda_{NP}^2}
\frac{U^d_{L33} U^u_{L32} U^\ell_{L33}}{V_{cb}} \right\vert^2 ~.
\eeq
(Here we have used the fact that $\sum_i |U_{L3i}^\nu|^2 = 1$.) The
addition of the NP operator thus has the effect of modifying the SM
prediction for $\Gamma(b \to c \tau {\bar\nu}_i)$ by an overall factor
that is lepton flavor non-universal. In fact, if the elements of the
charged-lepton mixing matrix obey the hierarchy suggested by GGL,
namely $|U^\ell_{L33}| \simeq 1$ and $|U^\ell_{L31}|^2 \ll
|U^\ell_{L32}|^2 \ll 1$, then $b \to c \tau {\bar\nu}_i$ is affected
by the NP, but $b \to c \mu {\bar\nu}_i$ and $b \to c e {\bar\nu}_i$
are basically unchanged from the SM.

We now have the simple prediction
\bea
\left[\frac{R(D)}{R(D^*)}\right]_{exp} & = & \left[\frac{R(D)}{R(D^*)}\right]_{SM} ~.
\label{prediction}
\eea
Using Eq.~(\ref{babarnew}), we have
\beq
\left[\frac{R(D)}{R(D^*)}\right]_{exp} = 1.33 \pm 0.24 ~~,~~~~
\left[\frac{R(D)}{R(D^*)}\right]_{SM} = 1.2 \pm 0.07 ~.
\eeq
So this model is consistent with experiment, but a careful measurement
of the double ratio can rule it out. The double ratio in the SM is
also likely to have less uncertainty from hadronic form
factors. Furthermore, all angular asymmetries, such as the $D^*$
polarization, forward-backward asymmetries, and the azimuthal angle
asymmetries including the triple products, will show no deviation from
the SM as these asymmetries probe non-SM operator structures.

If the ratios $R(D^{(*)})$ are defined with respect to the $B \to
D^{(*)} \mu \nu $ decay mode, we can also write
\beq
\left[\frac{R(D^*)_{exp}}{R(D^*)_{SM}}\right] = \left[\frac{R(D)_{exp}}{R(D)_{SM}}\right]
= \frac{\left[ 1+ 2 \, {\rm Re}(V_L^{cb\tau\nu_\tau}) + |V_L^{cb\tau}|^2 \right]}
      {\left[ 1+ 2 \, {\rm Re}(V_L^{cb\mu\nu_\mu}) + |V_L^{cb\mu}|^2 \right]} ~.
\eeq
Again assuming a hierarchy in the mixing matrix, to leading order we have
\bea
\left[\frac{R(D^*)_{exp}}{R(D^*)_{SM}}\right] & = & \left[\frac{R(D)_{exp}}{R(D)_{SM}}\right]
 \approx  \left[ 1+ 8 \frac{ g_2^2}{g^2}\frac{M_W^2}{\Lambda^2} \frac{ U^u_{L32}}{V_{cb}} \right] ~.
\eea
Averaging $\left[R(D^*)_{exp}/R(D^*)_{SM}\right]$ and
$\left[R(D)_{exp}/R(D)_{SM}\right]$, we get
\bea
 8 \frac{ g_2^2}{ g^2}\frac{M_W^2}{\Lambda^2} \frac{ U^u_{L32}}{V_{cb}} & \approx & 0.4 ~.
\eea
Taking $g/2 \lsim g_2 \lsim g$ and $ \Lambda \sim 10 M_W$, this gives
$0.8 \gsim U^u_{L32} \gsim \lambda$.

There have been numerous analyses examining NP explanations of the
$R(D^{(*)})$ measurements \cite{RDtheory,RDNP}.  Above, in the context
of $R_K$, we noted that, assuming the scale of NP is much larger than
the weak scale, all NP operators must be invariant under the full
$SU(3)_C \times SU(2)_L \times U(1)_Y$ gauge group. This same argument
applies also to NP proposed to explain $R(D^{(*)})$.  Such
considerations were applied to the semileptonic $ b \to c$ transitions
in Ref.~\cite{Feger:2010qc}, but they could have important implication
for the various NP explanations of the $R(D^{(*)})$ puzzle.

To sum up, the recent measurement of $R_K \equiv {\cal B}(B^+ \to K^+
\mu^+ \mu^-)/{\cal B}(B^+ \to K^+ e^+ e^-)$ by the LHCb Collaboration
differs from the SM prediction of $R_K = 1$ by $2.6\sigma$. And the
BaBar Collaboration has measured $R(D^{(*)}) \equiv {\cal B}({\bar B}
\to D^{(*)+} \tau^- {\bar\nu}_\tau)/$ ${\cal B} ({\bar B} \to D^{(*)+}
\ell^- {\bar\nu}_\ell)$ ($\ell = e,\mu$), finding discrepancies with
the SM of 2$\sigma$ ($R(D)$) and 2.7$\sigma$ ($R(D^*)$). The $R_K$ and
$R(D^{(*)})$ puzzles exhibit lepton flavor non-universality, and therefore
hint at new physics (NP).

Recently, Glashow, Guadagnoli and Lane (GGL) proposed an explanation
of the $R_K$ puzzle. They assume that the NP couples preferentially to
the third generation, and generates the neutral-current operator
$({\bar b}'_L \gamma_\mu b'_L) ({\bar \tau}'_L \gamma^\mu \tau'_L)$,
where the primed fields denote states in the gauge basis. When one
transforms to the mass basis, one obtains operators that give rise to
decays that violate lepton universality (and lepton flavor
conservation).

It is known that, assuming the scale of NP is much larger than the
weak scale, all NP operators must be made invariant under the full
$SU(3)_C \times SU(2)_L \times U(1)_Y$ gauge group. In this Letter, we
find that, when this is applied to the GGL operator, there are two
types of fully gauge-invariant NP operators that are possible. And one
of these contains both neutral-current and charged-current
interactions. While GGL has shown that the neutral-current piece of
this NP operator can explain the $R_K$ puzzle, we demonstrate that the
charged-current piece can simultaneously explain the $R(D^{(*)})$
puzzle. We also show that this model makes a prediction for the double
ratio $R(D)/R(D^*)$, so that it can be ruled out with a more precise
measurement of this quantity.

\bigskip
\noindent
{\bf Acknowledgments}:
This work was financially supported by the IPP (BB), by NSERC of Canada
(BB, DL), and  by the National Science Foundation (AD, SS) under Grant No.\
NSF PHY-1414345. We thank Olivier Mattelaer for useful discussions about
MadGraph 5.

\end{document}